\documentclass[aps,pra,superscriptaddress,amsmath,amssymb,preprintnumbers,twocolumn,showpacs]{revtex4}
\usepackage{amssymb}
\usepackage{color}

\newcommand{\Ket}[1]{\left\vert#1\right\rangle}
\newcommand{\BraKet}[2]{\left\langle#1\right\vert\left.#2\right\rangle}
\newcommand{\KetBra}[2]{\left\vert#1\right\rangle\left\langle#2\right\vert}
\newcommand{\Projector}[1]{\KetBra{#1}{#1}}
\newcommand{\MatrixEl}[3]{\left\langle#1\right\vert #2 \left\vert#3\right\rangle}

\newcommand{\belong}{\epsilon}
\newcommand{\ho}{\hat{H}_0}
\newcommand{\hv}{\hat{H}_v}
\newcommand{\cpc}{\gamma}
\newcommand{\cpf}{\hat{\Omega}}
\newcommand{\hvxpgen}[1]{\cpc#1\hat{\sigma}_{+}+h.c.}
\newcommand{\hvxp}{\hvxpgen{\cpf}}
\newcommand{\preservedset}{I(\tau)}
\newcommand{\ptarget}{\hat{P}_{d}}
\newcommand{\eqr}[1]{eq.(\ref{#1})}

\begin{document}


\title{Distilling angular momentum nonclassical states in trapped ions}


\author{B. Militello}
\author{A. Messina}
\affiliation{INFM, MIUR, and Dipartimento di Scienze Fisiche ed
Astronomiche dell'Universit\`{a} di Palermo, Via Archirafi, 36,
I-90123 Palermo. Italy}


\begin{abstract}
  In the spirit of Quantum Non-Demolition Measurements, we show that exploiting
  suitable vibronic couplings and repeatedly measuring
  the atomic population of a confined ion, it is possible to
  distill center of mass vibrational states with well defined square of angular
  momentum or, alternatively, angular momentum projection Schr\"odinger cat
  states.
\end{abstract}


\maketitle


\section{Introduction}

The harmonic oscillator is the very archetypical and basic system
both in quantum and classical physics. In fact, not only it
provides a good description of a physical system \lq moving\rq\ in
the nearby of a minimum of its potential energy, but in addition
it describes the independent modes of the electromagnetic field,
throwing a bridge between (quantum) mechanics and (quantum)
electrodynamics.

Today it is possible to confine an ion into a Paul trap in such a
way that its center of mass behaves as a three-dimensional
harmonic oscillator~\cite{TrappedIonReview_NistInnsbruck,Ghosh}. A
trapped ion, besides these vibrational degrees of freedom,
possesses fermionic dynamical variables too, related to its
internal state.

In such a physical context a wide variety of schemes for
generating both classical and nonclassical states have been
proposed and practically
realized~\cite{TrappedIonReview_NistInnsbruck,Ghosh,
TrappedIonReview_Vogel, TrappedIonReview_Nist}. We mention here
coherent, Fock, and squeezed states~\cite{Generation_Fock_Nist},
and Schr\"odinger Cat states involving entanglement between the
internal degrees of freedom (the atom) and the center of mass
quasi classical motion (two motion coherent states playing the
role of the two cat's conditions \lq alive\rq\ and \lq dead\rq\
are considered)~\cite{Generation_Cat_Nist}. Moreover coherent
superpositions of {\em macroscopically distinguishable} states,
often referred to as Schr\"odinger Cat states too, have been
reported. In particular it has been proposed the generation of the
superposition of two harmonic oscillator energy eigenstates
characterized by opposite angular momentum projections
corresponding to clockwise and counterclockwise center of mass
motions~\cite{Sabrina_Generation,Sabrina_AngularMomentum}. Such a
generation scheme, assuming that the trapped particle is
completely \lq isolated\rq, involves a long time interaction.
Moreover it requires a high degree of accuracy in chronological
control of the experiment, in the sense that the system is
described by the \lq Cat\rq\ state during a very small time
interval. Such critical points may be overcome exploiting
distillation processes~\cite{qpp} largely used in applications for
quantum technology~\cite{q_tech}, i.e. quantum computation,
quantum information and quantum teleportation.

Generally speaking, distilling a state means nothing but {\em
extracting} it from the initial state of the system, providing in
some sense the realization of projection operators. A wide class
of distillation processes are based on the idea that a physical
system, the {\em Slave} ($S$), in interaction with a repeatedly
measured one, the {\em Master} ($M$), undergoes a non-unitary
evolution provoking the decay of most of the quantum states (the
\lq residual\rq) in favor of few preserved ones (the \lq
distillate\rq)~\cite{Nakazato_PRL}. The relevant selection rule is
related to both the specific Master-Slave interaction and the
Master measurement results. As a very famous example we mention
Quantum Non-Demolition Measurements (QND)~\cite{Braginski_QND,
Walls_QND}, largely used for instance in trapped ions for
generating Fock states~\cite{Vogel_QND, Davidovich_QND,
Sabrina_QND}

In this paper we give the general sketch of a distillation
strategy developing the analysis in the specific framework of a
wide class of vibronic couplings which are realizable in the
context of trapped ions. The statement of this general approach is
then exploited in different applications. After revisiting the QND
scheme, the possibility of generating a superposition of those
Fock states whose quantum numbers correspond to \lq perfect
squares\rq\ is proved. The crucial result of this paper is
presented in the third section wherein the previously mentioned
generation of angular momentum Schr\"odinger Cat states in
two-dimensional traps is reported. The class of distilled \lq
Cats\rq\ turns out to be more general than that generated by the
method in ref~\cite{Sabrina_Generation}. Indeed, in our case, a
higher level of controllability of the quantum phase between the
two terms of the superposition is obtained. The efficiency and the
fidelity of the method are discussed and shown to be good enough.
In the fourth section the distillation procedure is applied to a
three-dimensional isotropically trapped ion, providing the
possibility of generating eigenstates of the square of the orbital
(i.e. related to the center of mass motion) angular momentum.
Finally, in the last section, some conclusive remarks are given.

\section{Distillation processes in trapped ions}

In a Paul trap a time dependent inhomogeneous (quadrupolar) field
is able to force a charged particle to move approximately as a
harmonic oscillator. Hence an ion confined in such a device
provides a compound system possessing both fermionic (electronic)
and bosonic (vibrational) degrees of freedom. The first ones
describe the internal state of the ion, i.e. the motion of the
electrons around the nucleus, and in most of the cases may be
represented as a two-level system. The other degrees of freedom
describe the oscillatory motion of the ion center of mass. The
relevant {\em unperturbed hamiltonian} is expressible as
($\hbar=1$)
\begin{equation}\label{TrappedIonUnperturbed}
  \ho=\sum_{i=x,y,z}\nu_i\hat{a}^{\dag}_i\hat{a}_i +
  \frac{\omega_0}{2}\hat{\sigma}_3
\end{equation}
where $\nu_i$ are the center of mass harmonic oscillator
frequencies, $\hat{a}_i$ ($\hat{a}^{\dag}_i$) the related
annihilation (creation) operators, $\omega_0$ is the Bohr
frequency between the two atomic levels considered, and
$\hat{\sigma}_3$ is the diagonal Pauli operator.

Acting upon the system through laser fields, it is possible to
implement a wide variety of vibronic couplings whose features
depend on the laser frequencies, wavelengths, polarizations and
strengths. Generally speaking, the $\ho$--interaction picture
hamiltonian model evaluated in the Rotating Wave Approximation
({\em RWA}) turns out to be time--independent and expressible
as~\cite{TrappedIonReview_NistInnsbruck,TrappedIonReview_Nist,
TrappedIonReview_Vogel}
\begin{equation}\label{TrappedIonVibronicCoupling}
  \hv=\hvxp
\end{equation}
where $\hat{\sigma}_{+}=\KetBra{+}{-}$
($\hat{\sigma}_{-}=\KetBra{-}{+}$) is the Pauli raising (lowering)
operator, $\Ket{\pm}$ being the internal ionic states, and $\cpc$
is a positive coupling constant related to laser intensities and
initial phases. The generic time--independent vibrational operator
$\cpf$ is a function of the annihilation and creation operators
$\{\hat{a}_i\}\cup\{\hat{a}^{\dag}_i\}$. Its specific form is
determined once the specific laser field configuration is
given~\cite{TrappedIonReview_Vogel}.

Assume the fermionic part of the compound system ($M$) is
initially in the state $\Ket{+}$, hence starting with the density
operator
\begin{equation}\label{TrappedIonInitialState}
  \hat{\rho}=\hat{\rho}_v\Projector{+}
\end{equation}
$\hat{\rho}_v$ being the initial vibrational state (the initial
state of $S$). Let the system evolve under the action of the
hamiltonian $\hv$ ($M$--$S$ interaction) for a time $\tau$, and
then measure the internal ionic state. Assume the system is found
in $\Ket{+}$, and then let again the system evolve in accordance
with $\hv$ for a time $\tau$, and measure the fermionic state
finding it in $\Ket{+}$, and so on $N$ times.

Under these assumptions, the system undergoes the non-unitary
evolution described by
\begin{equation}\label{NonUnitaryEvolution}
  \hat{W}_{+}^{(N)}(\tau)\equiv\aleph_N\left[\Projector{+}e^{-i\hv\tau}\right]^N\Projector{+}
\end{equation}
which may be cast in the form
\begin{equation}\label{NonUnitaryEvolution_Expression}
  \hat{W}_{+}^{(N)}(\tau)=\aleph_N\left[\hat{V}(\tau)\right]^N\Projector{+}
\end{equation}
with
\begin{equation}\label{NonUnitaryOperator}
  \hat{V}(\tau)\equiv\MatrixEl{+}{e^{-i\hv\tau}}{+}.
\end{equation}
and $\aleph_N=\left[\prod_{k=1}^{N}\sqrt{\wp_k}\right]^{-1}$,
$\wp_k$ being the probability of finding the ion  into the state
$\Ket{+}$ at the $k$th measurement step.

It is straightforward to prove that:
\begin{equation}\label{VibronicCouplingSquare}
  \hv^2=\cpc^2\left(\cpf^{\dag}\cpf\Projector{-}+\cpf\cpf^{\dag}\Projector{+}\right)
\end{equation}
\begin{equation}\label{VibronicCouplingEvenPower}
  \hv^{2n}=\cpc^{2n}\left[
    \left( \cpf^{\dag}\cpf \right)^n \Projector{-}+
    \left( \cpf\cpf^{\dag} \right)^n \Projector{+}
  \right]
\end{equation}
from which it immediately follows,
\begin{equation}\label{VibronicCouplingReduction}
  \begin{cases}
    \MatrixEl{+}{\hv^{2n+1}}{+}&=\hat{0}\cr
    \MatrixEl{+}{\hv^{2n}}{+}&=\cpc^{2n}\left(\cpf\cpf^{\dag}\right)^n\cr
  \end{cases}
\end{equation}

On the basis of these results, the unitary evolution operator
associated to $\hv$,
\begin{equation}\label{UnitaryEvolution}
  e^{-i\hv\tau}=
  \sum_{n=0}^{\infty} (-1)^n \frac {\hv^{2n}\;\tau^{2n}} {(2n)!}-i
  \sum_{n=0}^{\infty} (-1)^n \frac {\hv^{2n+1}\;\tau^{2n+1}}
  {(2n+1)!},
\end{equation}
restricted to the $\Ket{+}$ Master state, furnishes
\begin{equation}\label{NonUnitaryOperatorVibronic}
  \hat{V}(\tau)=\cos\left(\cpc\tau\sqrt{\cpf\cpf^{\dag}}\right)
\end{equation}

Such a non-unitary operator is a {\em real function} of the
hermitian non-negative operator $\cpf\cpf^{\dag}$, and hence is
hermitian too, and its eigensolution problem is strictly related
to that of $\cpf\cpf^{\dag}$ (same eigenstates, different
eigenvalues).

Let us denote by $\Ket{\omega_k}$ and $\omega_k$ the eigenstates
and the eigenvalues of $\cpf\cpf^{\dag}$ respectively, and
accordingly give the spectral decomposition
$\cpf\cpf^{\dag}=\sum_k \omega_k\Projector{\omega_k}$. Choose the
interaction time $\tau$ in such a way that for some $k$ it results
$|\cos(\cpc\tau\sqrt{\omega_k})|=1$.

Thus, for large enough $N$ it turns out
\begin{equation}\label{CosinusNthPower}
  \begin{cases}
    \cos^N\left(\cpc\tau\sqrt{\omega_k}\right)=(-1)^{l_k\cdot N}&k\ \belong\ \preservedset\cr
    \cos^N\left(\cpc\tau\sqrt{\omega_k}\right)\approx 0&k\ \not\belong\ \preservedset\cr
  \end{cases}
\end{equation}
where
\begin{equation}\label{DefinitionOf_I_Of_tau}
\preservedset\equiv\left\{k:\ \cpc\tau\sqrt{\omega_k}=l_k\pi,\
l_k\belong \mathbb{Z} \right\}
\end{equation}

Finally one has the following non-unitary action on the compound
system:
\begin{equation}\label{NonUnitaryEvolution_Projector}
  \hat{W}_{+}^{(N)}(\tau)\approx\aleph_N\Projector{+}\sum_{k\belong \preservedset}
  (-1)^{l_k\cdot N}\Projector{\omega_k}
\end{equation}
which is equivalent to the following non-unitary action on the
Slave:
\begin{equation}\label{NonUnitaryEvolution_Projector2}
  \hat{V}^{N}(\tau)\approx \sum_{k\belong \preservedset}
  (-1)^{l_k\cdot N}\Projector{\omega_k}\equiv
  e^{-i\ptarget\hat{G}\ptarget}\ptarget
\end{equation}
$\hat{G}$ being a suitable hermitian operator whose restriction
$\ptarget\hat{G}\ptarget$ to the subspace $\ptarget=\sum_{k\belong
\preservedset}\Projector{\omega_k}$ generates a unitary
transformation into the target subspace. When all $l_k\cdot N$ are
even it turns out $e^{-i\ptarget\hat{G}\ptarget}=\hat{1}$, and
hence $\hat{V}^{N}(\tau)\approx\ptarget$.

It is worth noting that in order to obtain the final result of the
distillation process we do not need detailed knowledge of the
dynamics induced by the hamiltonian in
\eqr{TrappedIonVibronicCoupling}, but just the diagonalization of
the positive hermitian operator $\cpf\cpf^{\dag}$. This
simplification shall reveal to be very effective and useful.

It deserves to be remarked that the distillation is a conditional
procedure in the sense that its success depends on $N$ {\em
stochastic events}. In other words, it is required that the
electronic system ($M$), at each measurement act, is always found
into the upper level $\Ket{+}$. Otherwise the procedure fails.
Moreover, the procedure is substantially the realization of a
projection operator, hence, for the process being successful, the
distilled states should be present in the initial vibrational
condition $\hat{\rho}_v$. Both these problematic aspects of the
method are solved considering that, as in the case of QND, the
joint probability of finding the Master system in its initial
state at each step ($\prod_{k=1}^{N}\wp_k$) tends, in the limit
$N\rightarrow\infty$, to the probability of finding the target
state (i.e. the \lq distillate\rq) into the initial Slave state
(see appendix \ref{App_Efficiency}), that is
\begin{equation}\label{Distillation_Efficiency}
\prod_{k=1}^{N}\wp_k\;\rightarrow\;Tr_S\{\hat{\rho}_v\ptarget\}
\end{equation}
being $Tr_S$ the trace operation over the Slave degrees of
freedom.

The quantity in \eqr{Distillation_Efficiency} expresses the
efficiency of the distillation process, that is the probability of
distillation success. Incidentally, \eqr{Distillation_Efficiency}
explains also the fact that QND may be used in trapped ions both
as a strategy for generating states and for measuring vibrational
state populations~\cite{Vogel_QND}.

\subsection{Quantum Non-Demolition Measurements of Single Fock States}

Let us consider as a specific example of this theory the standard
Quantum Non-Demolition Measurements.

Consider a laser field directed for instance along $x$, and tuned
to the atomic Bohr frequency $\omega_0$. The relevant vibronic
coupling has the form~\cite{TrappedIonReview_Vogel}
\begin{equation}\label{TrappedIonVibronicCouplingQND}
  \hv=\hvxpgen{f(\hat{a}^{\dag}_x\hat{a}_x, \eta_x)}
\end{equation}
where
\begin{eqnarray}\label{TrappedIonDef_Of_f}
  \nonumber
  f(\hat{a}^{\dag}_x\hat{a}_x,
  \eta_x)&\equiv&
  e^{-\frac{\eta_x^2}{2}}\sum_{l=0}^{\infty}\frac{(i\eta_x)^{2l}}{(l!)^2}\hat{a}_x^{\dag\
  l}\hat{a}_x\\
  &=& \sum_{n=0}^{\infty} L^{(0)}_{n}(\eta_x^2)\Projector{n}
\end{eqnarray}
where $L^{(0)}_{n}(\eta_x^2)$ is a Laguerre polynomial $\eta_x$
being the Lamb--Dicke parameter defined as the ratio between the
dimension of the oscillations of the ion center of mass in its
ground vibrational state, and the laser field wavelength.
Referring to \eqr{TrappedIonVibronicCoupling} we have
$\cpf=\cpf^{\dag}=f(\hat{a}^{\dag}_x\hat{a}_x, \eta_x)$, so that
it results
\begin{equation}\label{NonUnitaryOperatorQND}
  \hat{V}(\tau)=\cos\left[\cpc\tau
  f(\hat{a}^{\dag}_x\hat{a}_x,\eta_x)\right]
\end{equation}
It is easy to see that when $\eta_x\ll 1$ the function $f(n,
\eta_x)$ approaches unity for any $n$, otherwise such a function
is strongly nonlinear, and it turns out that different values of
$n$ give rise to incommensurable values of $f$. Therefore it is
possible to choose the interaction time $\tau$ in such a way that
$\cpc\tau f(\bar{n}_x, \eta_x)=2\pi$, and $\cpc\tau f(\bar{n}_x,
\eta_x)\not=l_{n}\pi$ for any $n\not=\bar{n}$, obtaining, for
large enough $N$,
\begin{equation}\label{CosinusNthPowerQND}
  \begin{cases}
    \cos^N\left[\cpc\tau f(\eta_x, \bar{n})\right]=1&\cr
    \cos^N\left[\cpc\tau f(\eta_x, n)\right] \approx 0&n\not=\bar{n}\cr
  \end{cases}
\end{equation}

This provides the decay of the Fock states $\Ket{n\not=\bar{n}}$
in favor of the selected state $\Ket{\bar{n}}$.

\subsection{Distillation of \lq Perfect-Square\rq\ Fock
states}

As another application consider the action of a laser field again
directed along $x$ but tuned to the first blue sideband, i.e.
tuned to the frequency $\omega_0+\nu_x$. In the Lamb--Dicke limit,
i.e. assuming $\eta_x\ll 1$, the interaction picture hamiltonian
model turns out to be approximately the Anti-Jaynes-Cummings
model:
\begin{equation}\label{TrappedIonVibronicCouplingAJC}
  \hv=\hvxpgen{\hat{a}^{\dag}_x}
\end{equation}
This means that $\cpf=\hat{a}^{\dag}_x$. Therefore, the relevant
$\hat{V}(\tau)$ is
\begin{equation}\label{NonUnitaryOperatorAJC}
  \hat{V}(\tau)=\cos\left(\cpc\tau\sqrt{\hat{a}^{\dag}\hat{a}}\right)
\end{equation}

It is straightforward to prove that, if $\cpc\tau=2\pi$ is chosen,
it results
\begin{equation}\label{CosinusNthPowerAJC}
  \begin{cases}
    \cos^N\left[\cpc\tau\sqrt{n})\right]= 1& \mbox{if $n$ is a square}\cr
    \cos^N\left[\cpc\tau\sqrt{n})\right] \approx 0& \mbox{otherwise}\cr
  \end{cases}
\end{equation}

Therefore the distillation procedure extracts the perfect square
Fock states provoking the decay of all non-perfect square number
states.

\section{Distilling Two-Dimensional Angular Momentum Schr\"odinger Cats}

Let us now go through one of the two main results of this paper.
It concerns the distillation of superpositions of vibrational
states corresponding to bidimensional trapped ion center of mass
motions characterized by well defined vibrational total excitation
number and {\em opposite} projections of the angular momentum. In
this sense we succeed in distilling angular momentum Schr\"odinger
Cat states. A scheme for generating such a kind of superpositions
has been already proposed~\cite{Sabrina_Generation}. In fact,
starting from a specific initial condition (a Fock state), and
subjecting the system to the action of a bilinear two-mode
Jaynes-Cummings-like hamiltonian model ($\propto
\hat{a}_x\hat{a}_y\hat{\sigma}_+ + h.c.$), at some specific
instants of time the ion center of mass is found in the mentioned
\lq Cat\rq. Nevertheless, such a procedure requires a very high
degree of accuracy in temporal control of the experiment. We
propose here to exploit the general distillation strategy to
extract the same state generated with the recalled scheme.
Moreover, the class of states we succeed in generating is wider
than the previously mentioned. In fact, the phase relation between
the two terms of the superposition is in our case very easy to
control.

Consider an isotropic bidimensional Paul trap, that is assume for
instance $\nu_x=\nu_y\equiv\nu$ and neglect the motion along $z$.
Act on the system simultaneously through two laser beams both
tuned to the second red sideband $\omega_0-2\nu$. Let the two
lasers be responsible for the same coupling strength. The relevant
interaction hamiltonian is~\cite{Sabrina_Generation}
\begin{equation}\label{TrappedIonVibronicCouplingAngMom2D}
  \hv=\hvxpgen{\left(\hat{a}_x^2+\hat{a}_y^2\right)}
\end{equation}

Observe now that the operator $\cpf\cpf^{\dag}$ may be cast in the
form
\begin{equation}\label{NonUnitaryOperatorAngularMomentum2D}
  \cpf\cpf^{\dag}=\sum_{j,k=x,y}\hat{a}_j^2\hat{a}_k^{\dag
  2}=\hat{N}_T^2+4\hat{N}_T-\hat{L}_z^2+4
\end{equation}
being
$\hat{N}_T=\hat{a}_x^{\dag}\hat{a}_x+\hat{a}_y^{\dag}\hat{a}_y$,
and
$\hat{L}_z=i\left[\hat{a}_x\hat{a}_y^{\dag}-\hat{a}_y\hat{a}_x^{\dag}\right]$.

On the basis of \eqr{NonUnitaryOperatorAngularMomentum2D} it turns
out that, suitably choosing $\cpc\tau$, the distillation process
in this case may preserve those states characterized by specific
excitation number ($n_T$) and angular momentum projection ($m$),
denoted by $\Ket{n_T, m}$, such that $\cpc\tau\sqrt{n_T^2+4
n_T-m^2+4}=l_m\pi$, $l_m$ being an integer. Observe now that both
$m$ and $-m$ eventually satisfy such a relation. Moreover, that in
general for different $n_T$ and/or absolute value of $m$ the
relevant square roots result to be {\em incommensurable}.
Therefore experimental details, like coupling strength coupling
($\cpc$), interaction time ($\tau$), etc., may be adjusted to
project the system into the bidimensional subspace generated by
$\{\Ket{n_T, \pm m}\}$, in accordance with \eqr{CosinusNthPower}.
Thus, if the initial condition is a state whose overlaps with
$\Ket{n_T, \pm m}$ have the same modulus, the distilling procedure
returns the $50\%$-superposition of such angular momentum
eigenstates.

As a very specific example consider the trapped ion prepared in
the Fock state possessing $n_T=4$ vibrational excitations and
directed along $x$, $\Ket{n_x=4, n_y=0}$.  Such a state is
expressible as a superposition of the states $\Ket{n_T=4, m}$ with
$m=0,\pm 2,\pm 4$. It is straightforward to calculate
$\BraKet{n_T=4,m=\pm 4}{n_x=4, n_y=0}=\frac{1}{4}$, and to verify
that the \lq square roots\rq\ corresponding to different absolute
values of $m$ are incommensurable ($2\sqrt{5}, 4\sqrt{2}, 6$ for
$m=\pm 4, \pm 2, 0$ respectively). Adjusting $\tau$ in such a way
that $2\sqrt{5}\cpc\tau=2\pi$ one easily distillates, up to a
global phase factor, the superposition
\begin{equation}\label{SchroedingerCatAngularMomentum2D}
  \Ket{\psi_{dist}}=\frac{1}{\sqrt{2}}\left[\Ket{n_T=4, m=4}+\Ket{n_T=4,
  m=-4}\right]
\end{equation}
which, as announced, is just an angular momentum Schr\"odinger cat
superposition.

Straightforwardly one obtains that the efficiency of the
distillation is $\frac{1}{4^2}+\frac{1}{4^2}=\frac{1}{8}$, and
that, after $N=5$ steps, the Schr\"odinger cat state is distilled
with a $5\%$-error.

In this specific example the relative phase factor between the two
states of the superposition turns out to be $1$. Nevertheless,
selecting a different direction of the initial motion, it is
possible to change the relative phase between the two states in
the initial condition and hence in the distilled superposition. In
particular, denoting by $\theta$ the angle between the direction
of the initial motion and the $x$ axis, it results that the
relative phase between the two states of the distilled
superposition is $2 n_T\theta$, giving
\begin{equation}\label{SchroedingerCatAngularMomentum2DTheta}
  \Ket{\psi_{dist}}=\frac{1}{\sqrt{2}}\left[\Ket{n_T=4, m=4}+e^{i8\theta}\Ket{n_T=4,
  m=-4}\right]
\end{equation}

To prove such an assertion it is enough to consider that both a
Fock state directed along the direction rotated of $\theta$ around
$z$ with respect to the $x$ axis, denoted by $\Ket{n_{\theta}}$,
and the angular momentum eigenstates of maximum projection
$\Ket{n_T, m=\pm n_T }$, are $SU(2)$ states\cite{Perelomov}
expressible as
\begin{equation}\label{SU2}
  \Ket{\mu,j=\frac{n_T}{2}}=
  \frac{1}{(1+|\mu|^2)^{\frac{n_T}{2}}}
  \sum_{k=0}^{n_T}
        \left(\begin{array}{l}
                n_T \\
                k
              \end{array}
        \right)^{\frac{1}{2}}
        \mu^k\Ket{n_T-k,k}
\end{equation}
For $\mu=\tan\theta$ one has \lq rotated Fock states\rq\
($\Ket{n_{\theta}}$  $=$ $\Ket{\mu=\tan\theta,j=\frac{n_T}{2}}$),
that is an eigenstate of the number operator associated to the
annihilation operator
$\hat{a}_{\theta}\equiv\frac{1}{\sqrt{2}}\left[\cos\theta\hat{a}_x+\sin\theta\hat{a}_y\right]$.
For $\mu=\pm i$ one has maximum projection angular momentum
eigenstates ($\Ket{n_T,m=\pm n_T}=$ $\Ket{\mu=\pm i,
j=\frac{n_T}{2}}$~\cite{Sabrina_AngularMomentum}, which may be
directly verified. Their overlap is easily appraisable (see
appendix \ref{App_SU2Overlap}) and it turns out
\begin{eqnarray}\label{ScalarProduct}
  \nonumber
  \BraKet{n_T,m=\pm n_T}{n_{\theta}}=\frac{1}{2^{\frac{n_T}{2}}}e^{\mp i n_T\theta}
\end{eqnarray}

Therefore, as anticipated, the relative phase between the two
angular momentum eigenstates in both the initial and the distilled
states is $\xi(\theta) = 2 n_T\theta$.

The efficiency ($\frac{1}{8}$) and the agreement ($5\%$ after
$N=5$ steps) are the same as in the particular case $\theta=0$.

\section{Distilling Three-Dimensional Angular Momentum Eigenstates}

The last application we present in this paper concerns the
possibility of generating tridimensional states of motion
characterized by well defined vibrational total excitation number
and square of angular momentum. To this end consider an isotropic
tridimensional Paul trap ($\nu_x=\nu_y=\nu_z\equiv \nu$) which may
be easily realized adding to the standard quadrupolar sinusoidally
time--dependent electric field, a quadrupolar static electric
field~\cite{Tosheck}.

The action of three orthogonally propagating lasers tuned to the
second red sideband produces the tridimensional generalization of
the coupling term in \eqr{TrappedIonVibronicCouplingAngMom2D},
i.e. the vibronic coupling in \eqr{TrappedIonVibronicCoupling}
with

\begin{equation}
  \cpf=\hat{a}_x^2+\hat{a}_y^2+\hat{a}_z^2
\end{equation}

Few algebraic manipulations give
\begin{equation}\label{NonUnitaryOperatorAngularMomentum3D}
  \cpf\cpf^{\dag}=\sum_{j,k=x,y,z}\hat{a}_j^2\hat{a}_k^{\dag
  2}=\hat{N}_T^2+5\hat{N}_T-\vec{L}^2+6
\end{equation}
being $\hat{N}_T=\sum_{j=x,y,z}\hat{a}_j^{\dag}\hat{a}_j$, and
$\vec{L}=\left(\hat{L}_x,\hat{L}_y,\hat{L}_z\right)$ with
$\hat{L}_l=i\left[\hat{a}_j\hat{a}_k^{\dag}-\hat{a}_k\hat{a}_j^{\dag}\right]$
for $(j,k,l)=(x,y,z)\;and\; cyclics$.

As in the bidimensional case, the incommensurability of the square
roots $\sqrt{n_T^2+5 n_T-l(l+1)+6}$ for different values of $n_T$
and $l$ ensures the possibility of extracting prefixed states.
Here $n_T$ and $l$ are the vibrational total excitation and the
angular momentum square quantum numbers. As a consequence, a
single value of square of angular momentum may be selected. On the
other hand, the circumstance that the eigenvalues of $\hat{L}_z$
(or of any other component) do not appear in the square root
implies no selection of its eigenstates. Therefore, in general,
the result of the distillation process is a generic linear
combination
\begin{equation}\label{SuperpositionAngularMomentum3D}
  \Ket{\psi_{dist}}=\sum_{m=-l}^{l}c_m\Ket{n_T, l, m}
\end{equation}
involving states possessing well defined (i.e. the same for all)
$\hat{N}_T$ and $\vec{L}^2$ but different angular momentum
projections.

As an example we mention that starting with the Fock state
$\Ket{n_x=2, n_y=n_z=0}$, and choosing $\tau$ such that
$\gamma\tau 2\sqrt{5}=2\pi$, it is possible to distillate the
state $\Ket{n_T=2, l=0, m=0}$ with an efficiency $\frac{1}{3}$.

In passing we observe that, since the target angular momentum is
$l=0$, the distilled subspace turns out to be one-dimensional
without ambiguity for the projection ($m=0$).

It is worth noting that angular momentum eigenstates are in
general strongly entangled states in the Fock basis. Hence, their
generation provides an effective strategy for tridimensional
entangled-states preparation. Accordingly to our example, consider
indeed the Fock-basis expansion of the distilled state,
\begin{equation}\label{EntangledState}
  \Ket{n_T=2, l=0, m=0}=\frac{1}{\sqrt{3}}\left[\Ket{2,0,0}+\Ket{0,2,0}+\Ket{0,0,2}\right]
\end{equation}

It turns out that the spherically symmetric angular momentum
eigenstate considered, $\Ket{n_T=2, l=m=0}$, is a strongly
entangled state involving the three orthogonal motions of the ion
center of mass, possessing the same structure as a
$W$-state~\cite{W_State}.

In this case, after $N=5$ steps one reaches the target state in
\eqr{EntangledState} with less than $4\%$ error.

The action of the two- and three-dimensional reported distillation
processes may be in principle combined in order to realize
tridimensional center of mass motion \lq Scr\"odinger cats\rq\
characterized by well defined vibrational excitations and square
of angular momentum, but involving states with opposite
projections of angular momentum.

\section{Discussion and conclusive remarks}

In this paper we have presented a new strategy for distilling
single harmonic oscillator nonclassical states having the form of
superpositions of macroscopically distinguishable angular momentum
eigenstates (namely angular momentum Schr\"odinger Cat states). We
moreover succeeded, in the framework of the same distillation
method, to synthesize states possessing well defined square of
angular momentum. The physical scenario wherein these results have
been presented is that of a single trapped ion -- whose center of
mass moving in the trapping potential is the harmonic oscillator
under scrutiny.

The general distillation strategy that we use, recalled in the
second section, exploits repeated measurements on a system in
interaction with the one where we want to extract some interesting
states. In fact, in such a situation, the latter system undergoes
a non-unitary evolution provoking the decay of the undesired
states, keeping in life only few states (the \lq distillate\rq).
Giving an explicit expression of the non-unitary operator is not
always a trivial job. Moreover, due to such a non-unitariness, in
some cases it is impossible to perform its spectral decomposition.
Nevertheless, in the framework of the spin-boson interaction, and
in particular in the context of trapped ions, we succeed in
providing a very manageable expression for the effective
non-unitary operator acting upon the bosonic system (the center of
mass motion) when the fermionic system (the atomic internal state)
is repeatedly measured. Such a very expressive and useful result,
given in \eqr{NonUnitaryOperatorVibronic}, guarantees the
possibility of obtaining the spectral decomposition of the
non-unitary operator acting upon the system of interest,
$\hat{V}(\tau)=\cos(\cpc\tau\sqrt{\cpf\cpf^{\dag}})$, tracing such
a problem back to the spectral decomposition of the positive
hermitian operator $\cpf\cpf^{\dag}$. Therefore we have at our
disposal a powerful tool to forecast and control (suitably setting
the coupling strength, $\cpc$, and the interaction time, $\tau$)
the result of the distillation process.

Exploiting the potentialities of \eqr{NonUnitaryOperatorVibronic},
in the third section we have found that in a bidimensional
isotropic trap a suitable two-mode vibronic coupling is
responsible for distilling superpositions of vibrational states
possessing opposite angular momentum projections. Moreover, as
reported in the fourth section, the analogous ({\it mutatis
mutandis}) interaction in an isotropic tridimensional Paul trap
renders it possible the extraction of center of mass motion states
possessing well defined excitation number and square of angular
momentum. In a specific mentioned example such a state turns out
to have the structure of a $W$-state.

As a conclusive remark we state that, since in general the number
of steps required to distillate the target subspace is very small
(in the considered examples $N=5$ to have $95\%$ fidelity) the
duration of the experiment turns out to be short enough to
legitimate neglecting decoherence.


\appendix

\section{Distillation Efficiency}\label{App_Efficiency}

In this appendix we prove the limit expressed by
\eqr{Distillation_Efficiency}. Assume the system prepared, for
simplicity, into the pure state
$\Ket{\psi_0}=\Ket{\varphi_0}\Ket{+}$, $\Ket{\varphi_0}$ being the
initial vibrational state. The probability of finding the Master
system (fermionic degrees of freedom) in $\Ket{+}$ after the
unitary $M$-$S$ interaction is $\wp_1$ $=$
$||\MatrixEl{+}{e^{-i\hv\tau}}{\psi_0}||^2$ $=$
$||\hat{V}(\tau)\Ket{\varphi_0}||^2$,
$||\cdot||^2\equiv|\BraKet{\cdot}{\cdot}|^2$ denoting the relevant
Hilbert space \lq norma\rq\ . The resulting \lq collapsed\rq\
state is
$\Ket{\psi_1}=\frac{1}{\sqrt{\wp_1}}\left[\hat{V}(\tau)\Ket{\varphi_0}\right]\Ket{+}$.
Immediately after the second step one obtains
$\wp_2=||\MatrixEl{+}{e^{-i\hv\tau}}{\psi_1}||^2=\frac{1}{\wp_1}||\left[\hat{V}^2(\tau)\Ket{\varphi_0}\right]||^2$,
and
$\Ket{\psi_2}=\frac{1}{\sqrt{\wp_1\wp_2}}\left[\hat{V}^2(\tau)\Ket{\varphi_0}\right]\Ket{+}$.

In general it is
\begin{equation}\label{ProbN}
  \wp_n=\frac{1}{\wp_0\wp_1\cdot\cdot\wp_{n-1}}||\hat{V}^{n}(\tau)\Ket{\varphi_0}||^2
\end{equation}

\begin{equation}\label{StateN}
  \Ket{\psi_n}=\frac{1}{\sqrt{\wp_0\wp_1\cdot\cdot\wp_{n}}}\hat{V}^{n}\Ket{\varphi_0}\Ket{+}
\end{equation}

From \eqr{ProbN} it follows
\begin{equation}\label{CalculatedEfficacyLemma}
  \prod_{k=1}^{N}\wp_k=\wp_N\prod_{k=1}^{N-1}\wp_k=||\hat{V}^{N}(\tau)\Ket{\varphi_0}||^2
\end{equation}

Considering the limit $N$ $\rightarrow$ $\infty$, we can take
advantage of \eqr{NonUnitaryEvolution_Projector2} easily obtaining
\begin{equation}\label{CalculatedEfficiency}
  \prod_{k=1}^{N}\wp_k\approx||\ptarget\Ket{\varphi_0}||^2
\end{equation}
which expresses the same content of \eqr{Distillation_Efficiency}.

\section{Overlap between $SU(2)$ states}\label{App_SU2Overlap}

In this appendix we give the general expression of the overlap
between two two-mode $SU(2)$ states, and calculate the phase
between the maximum projection angular momentum eigenstates and
the rotated Fock states.

From the definition in \eqr{SU2}, the overlap is easily calculated
as
\begin{equation}\label{SU2OverlapFormula}
  \BraKet{\mu_1,j_1}{\mu_2,j_2}=\frac{(1+\mu_1^*\mu_2)^{2j}}{(1+|\mu_1|^2)^{j_1}(1+|\mu_2|^2)^{j_2}}\delta_{j_1,j_2}
\end{equation}

In the case $\mu_1=\pm i$ and $\mu_2=\tan\theta$ one obtains
\begin{eqnarray}\label{SU2OverlapFormula_Theta_I}
  \nonumber
  &&\BraKet{\mu_1=\pm
  i,j}{\mu_2=\tan\theta,j}=\frac{(1\mp\tan\theta)^{2j}}{2^j(1+\tan^2\theta)^j}\\
  &&=\frac{\left(\sqrt{1+\tan^2\theta}\ \ e^{\mp
  i\theta}\right)^{2j}}{2^j(1+\tan^2\theta)^j}=\frac{e^{\mp i 2 j \theta}}{2^j}
\end{eqnarray}

Therefore, the phase difference between the coefficients of the
$SU(2)$ states with $\mu=+i$ and $\mu=-i$ contained in
$\Ket{\mu=\tan\theta,j}$ results to be
$4j\theta=4\frac{n_T}{2}\theta=2 n_T\theta$.

\end{document}